\documentclass[prl,amsmath,amssymb]{revtex4}
\usepackage{times}
\usepackage{graphicx}
\usepackage{dcolumn}
\usepackage{bm}
\begin{document}
\title{Replacing energy by von Neumann entropy in  quantum phase transitions}
\author{Angela Kopp, Xun Jia, and Sudip Chakravarty$^{*}$}
\affiliation{Department of Physics and Astronomy, University of
California Los Angeles, Los Angeles, CA 90095-1547}
\date{\today}
\begin{abstract}
{\bf In the thermodynamic limit two distinct states of matter cannot be analytic continuations of each other.  Classical phase transitions are characterized by non-analyticities of the free energy. \cite{Yang:1952} For quantum phase transitions \cite{Sachdev:1999} (QPTs) the ground state energy often assumes the role of the free energy.  But in a number of important cases this criterion fails to predict a QPT, such as the three-dimensional metal-insulator transition of non-interacting electrons in a random potential (Anderson localization) \cite{Edwards:1971}.  It is therefore essential that  we find alternative criteria that can track fundamental changes in the internal correlations of the ground state wavefunction.  Here we propose that QPTs are generally accompanied by non-analyticities of the von Neumann (entanglement) entropy. \cite{Bennett:1996}  In particular, the entropy is non-analytic at the Anderson transition \cite{Abrahams:1979}, where it exhibits unusual fractal scaling.  We also examine  two dissipative quantum systems  of considerable interest to the study of decoherence and find that non-analyticities occur if and only if the system undergoes a  QPT. \cite{Chakravarty:1982}}
\end{abstract}
\email[Correspondence and requests for materials should be addressed to S.C.  ] { sudip@physics.ucla.edu}

\maketitle

Recently there has been intense interest in using entanglement to study  quantum phase transitions. \cite{Osterloch:2002,Osborne:2002,Refael:2004,Vidal:2003} Since entanglement is a unique quantum phenomenon which characterizes wave functions that cannot be factored  into single particle states, one expects that it should play a role in generating the correlations that exist on all length scales at a quantum critical point. But a state can be entangled without being critical---consider, for instance, the singlet state of two spin-$1/2$ particles. So it is not mere entanglement which concerns us here, but also the criticality of infinitely many interacting degrees of freedom.

In an important paper on black hole entropy, Bekenstein \cite{Bekenstein:1973} demonstrated the usefulness of information entropy. The concept is readily applicable to a quantum mechanical ground state. If we consider the the entire wave function, the conventional entropy is of course zero (assuming the state is non-degenerate), and all we can study  is the analyticity of the ground state energy. In those cases where this is a smooth analytic function of a tuning parameter that drives a QPT, we get no useful information. Yet in many cases we know that the wave function represents special internal correlations, as in Laughlin's quantum Hall wave function, \cite{Laughlin:1983} or in the Bijl-Dingle-Jastrow-Feynman wave function \cite{Feenberg:1969}  describing the superfluid state of bosons. How can we quantify such correlations, in particular when they change across a transition?

Let us divide a system into two pieces and calculate the entanglement between them. In the ground state $|\Psi\rangle$, the density matrix of the full system is $\rho=|\Psi\rangle\langle\Psi|$. Now partition the system into two parts $A$ and $B$, where $A$ denotes the subsystem of interest and $B$ the environment whose details are of no interest. We construct the reduced density matrix $\rho_{A}$ by tracing over the environmental degrees of freedom, akin to integrating out the microstates consistent with a given set of  macroscopic thermodynamic variables.  The von Neumann entropy, defined by $S=-\mathrm{Tr (\rho_{A} \ln \rho_{A})}$, provides a measure of bipartite entanglement and therefore contains information about the quantum correlations present in the ground state. Note that  $S=-\mathrm{Tr}(\rho_A \ln \rho_A)=-\mathrm{Tr}(\rho_B \ln \rho_B)$, where the reduced density matrix $\rho_B$ is obtained by tracing over the degrees of freedom in subsystem $A$. 

We consider three important models. In addition to demonstrating the use of von Neumann entropy, these models illuminate the role of disorder and dissipation in entanglement, {\it inter alia} in quantum computation. The two dissipative systems that we discuss have been  recently addressed from this point of view, \cite{Stauber:2006} but the results are incorrect.  Our first example, however, is  Anderson localization in three dimensions ($d=3$), which has been studied extensively and is known to have a quantum critical point.  A critical value of disorder separates a system where all states are localized from one where some states are extended. At the critical point the wave function exhibits a fractal character. \cite{Wegner:1980} 

The relevant Hamiltonian for a $d$-dimensional lattice, expressed in terms of fermionic creation ($c_{i}^{\dagger}$) and annihilation ($c_{i}$) operators,  is---suppressing spin indices, irrelevant in this case--- 
\begin{equation}
H= \sum_{i}E_{i}c^{\dagger}_{i}c_{i} - t \sum_{\langle i,j \rangle}\left(c^{\dagger}_{i}c_{j}+\text{h. c.}\right),
\end{equation}
where the sum in the second term is over distinct nearest-neighbor pairs, and the site energies $E_{i}$ are chosen from a uniform random distribution between $-W/2$ and $W/2$. In what follows, we shall choose the unit of energy such that $t=1$. Since this is a one-electron problem, the single-site density matrix in the state $\alpha$ is
\begin{equation}
\rho_{i}^{\alpha}=|\psi_{i}^{\alpha}|^{2}(|1\rangle \langle 1|)_{i}+(1-|\psi_{i}^{\alpha}|^{2})(|0\rangle \langle 0|)_{i}
\end{equation}
where $\psi_{i}^{\alpha}$ is the amplitude of an energy eigenstate $\alpha$ at site $i$. The ket $|0\rangle$ ($|1\rangle$) corresponds to the site being unoccupied (occupied). The von Neumann entropy, averaged over all the eigenstates $\cal A$ and the number of sites $N$, is
\begin{equation}
S = -\frac{1}{{\cal A} N}\sum_{n=1}^{N}\sum_{\alpha=1}^{\cal A}\left(|\psi_{i}^{\alpha}|^{2}\ln |\psi_{i}^{\alpha}|^{2}+(1- |\psi_{i}^{\alpha}|^{2})\ln (1- |\psi_{i}^{\alpha}|^{2})\right)
\end{equation}
For a random system $S$ must be further averaged over realizations of the random potentials.  The resulting quantity, denoted by 
$\overline S$, measures the correlation between the amplitude at a given site and the rest---local entanglement. \cite{Zanardi:2002}
\begin{figure}[htbp]
\begin{center}
\includegraphics[scale=1.0]{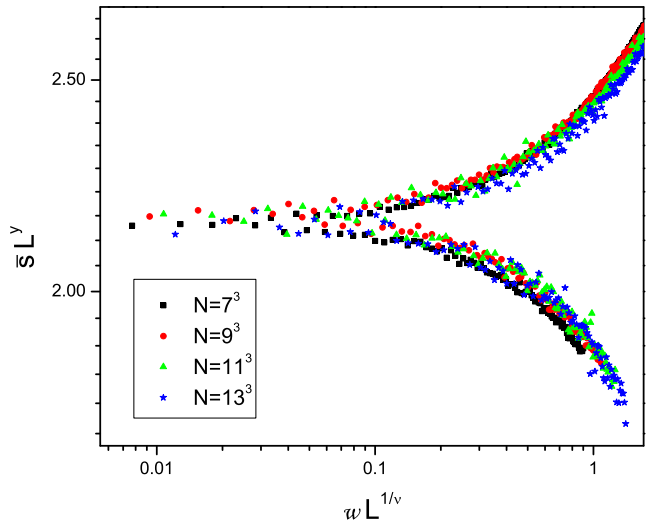}
\caption{The data collapse of the calculated von Neumann entropy of the $d=3$ Anderson localization problem. The exponent $\nu$ and $W_{c}$ were set to the known values 1.35  and 16.5, respectively. The exponent $y$ was then determined to be $2.6$. The system sizes are indicated in the inset. The systems of sizes $7^{3}$ and $9^{3}$ were averaged over 10 different realizations of disorder, while the system of size $11^{3}$ was averaged over 5 realizations. For the $13^{3}$ lattice, it did not seem necessary to average over multiple realizations.}
\label{fig:entropy}
\end{center}
\end{figure}

We compute $\overline S$ by diagonalizing the Hamiltonian for finite-sized systems in $d=3$ and then attempt to collapse the data using the finite size scaling hypothesis for the singular part of the entropy: \cite{Calabrese:2004}
\begin{equation}
{\overline S}_{\text{sing}} = L^{-y}f_{\pm}(L^{1/\nu}w),
\end{equation}
where $L=N^{1/3}$ is the length of the system, $w=|W-W_{c}|/W_{c}$, and $W_{c}$ is the critical disorder at which Anderson localization occurs. The critical exponent $\nu$ is the exponent for the localization length, the value of which is known to be approximately $1.35$; $W_{c}$ is known to be approximately 16.5. \cite{Hofstetter:1994}The universal scaling functions $f_{\pm}$ refer to $W$ greater or less than $W_{c}$. It has been conjectured that for dimensions $d>2$, the exponent $y=(d-1)$. \cite{Calabrese:2004} Examining the data collapse in Fig.~\ref{fig:entropy}, we find instead that the exponent $y=2.6$, a fractal dimension less than three.
\begin{figure}[htbp]
\begin{center}
\includegraphics[scale=1.0]{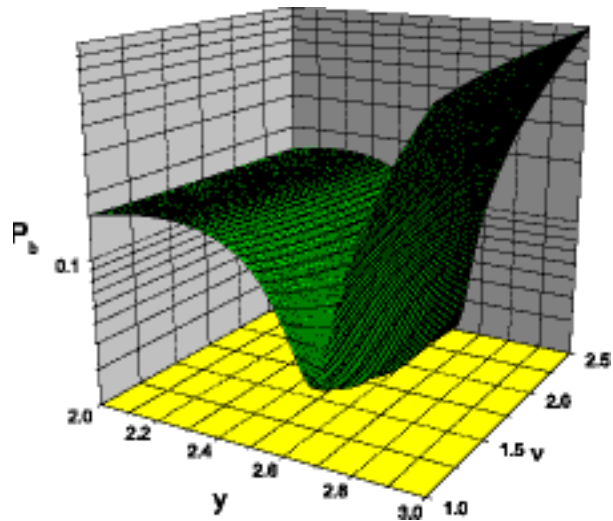}
\caption{The minimum of $P_{b}$ corresponds to the best data collapse. From the figure, we see that $y=2.6$ and $\nu=1.35$ are good choices, although the minimum with respect to $\nu$ is very shallow. }
\label{fig:collapse}
\end{center}
\end{figure}
This is similar to the fractal scaling of the participation ratio. \cite{Wegner:1980,Parshin:1999} The origin of the fractal dimension is the fractal nature of the wave function at the mobility edge separating the delocalized states from the localized states. Of course, data collapse without an objective criterion can be treacherous. An interesting criterion used to justify the goodness of collapse is shown in Fig.~\ref{fig:collapse} .  The quantitative measure is obtained from the minimization of a specially defined function of the residuals called $P_{b}$, \cite{Bhattacharjee:2001} which satisfies $P_{b}\ge P_{b}|_{{\mathrm{abs\: min}}}=0$.

\begin{figure}[htbp]
\begin{center}
\includegraphics[scale=1.0]{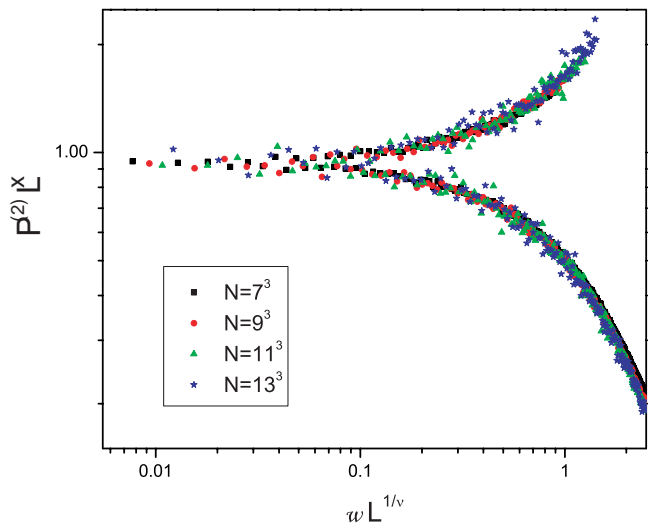}
\caption{The data collapse, as in Fig.~\ref{fig:entropy}, of the participation ratio of the $d=3$ Anderson localization problem. The exponent $\nu$ and $W_{c}$ were set to the known values 1.35  and 16.5, respectively. The exponent $x$ was then determined to be $1.4$.}
\label{fig:PR}
\end{center}
\end{figure}

The fractal dimension of the von Neumann entropy can be confirmed from another measure of entanglement called the linear entropy, \cite{Zurek:1993} which is $S_{L}=1-\mathrm{Tr \rho_{A}^{2}}$. For the present problem, it is easy to show that the participation ratio $P^{(2)}=\frac{1}{\cal A}\sum_{i,\alpha}|\psi_{i}^{\alpha}|^{4}=1-S_{L}/2$. This is to be averaged over the realizations of the random potential, if necessary.
In the extreme localized case, only one site participates and $S_{L}=0$. In the contrasting limit of fully extended state $S_{L}=2-2/N\to 2$, when $N\to \infty$. The participation ratio, hence $S_{L}$, exhibits scaling at the metal-insulator transition: $P^{(2)}=L^{-x} g_{\pm}(L^{1/\nu}w)$, where $g_{\pm}$ is another universal function. As shown in Fig.~\ref{fig:PR}, the fractal dimension $x=1.4$, consistent with the results known in the literature. 

The general model of a two-level system (a qubit) coupled to a dissipative heat bath is relevant to a number of interesting physical problems. \cite{Leggett:1987}  We will consider a specific formulation, the spin-boson model, \cite{Blume:1970} in which the two-level system is represented by a spin-1/2 degree of freedom and the bath is a collection of (bosonic) harmonic oscillators.  The Hamiltonian is
\begin{equation}
H_{\mathrm{sb}}=-\frac{1}{2} \, \Delta \, \sigma_x + H_{\rm osc} + \frac{1}{2} \, \sigma_z\sum_n \lambda_n x_n
\label{hamiltonian}
\end{equation}
where $\sigma_x$ and $\sigma_z$ are Pauli matrices and $H_{\rm osc}$ represents the Hamiltonian of an infinite number of harmonic oscillators.  The last term couples the $z$-component of the spin to the coordinates $\{x_n\}$ of the oscillators.  The coupling constants $\{\lambda_n\}$, together with the masses $\{m_n\}$ and frequencies $\{\omega_n\}$, determine the spectral density of the heat bath, which is given by
\begin{equation}
J(\omega)=\frac{\pi}{2} \sum_n \frac{\lambda_n^2}{m_n \omega_n} \, \delta(\omega-\omega_n)
\label{spectraldens}
\end{equation}
For an Ohmic bath, we can take $J(\omega)=2 \pi \alpha \omega$ for $\omega < \omega_{c}$ and $J(\omega)=0$ for $\omega \geq \omega_{c}$.

\begin{figure}[htbp]
\begin{center}
\includegraphics[scale=0.5]{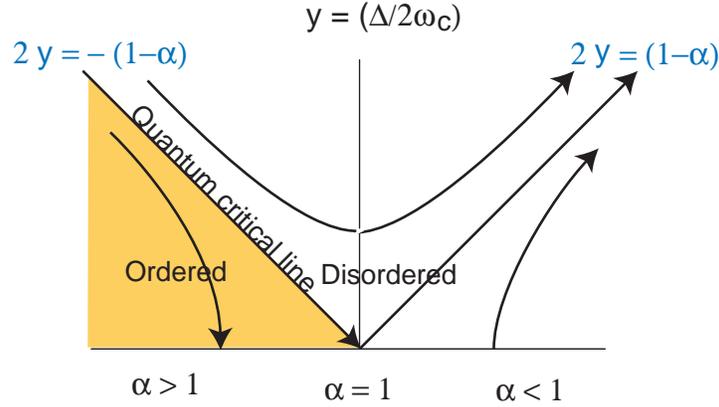}
\caption{The $T=0$ phase diagram of a two-level system coupled to an Ohmic heat bath and the corresponding renormalization group flows in the regime $|(1-\alpha)|\ll 1$ and $y\ll 1$. The renormalization group trajectories point along increasing values of the flow parameter $l=\ln(\tau \omega_c)$, where $\tau$ is imaginary time. The separatrix $2y=-(1-\alpha)$ separates a broken symmetry phase ($\langle \sigma_{z}\rangle\ne 0$) from the disordered phase ($\langle \sigma_{z}\rangle=0$). At this critical line the magnetization is discontinuous even though the correlation time (imaginary) diverges.  At $\alpha=1/2$ the dynamics of the two-level system changes from an overdamped to an underdamped state, but there is no thermodynamic phase transition at this point.}
\label{fig:phase}
\end{center}
\end{figure}

At zero temperature, this model has a quantum critical line separating a broken-symmetry phase with $\langle \sigma_z \rangle =M_{0}\neq 0$ from a disordered phase with $\langle \sigma_z \rangle = 0$ (see Fig.~\ref{fig:phase}). \cite{Chakravarty:1982}  As with any broken-symmetry phase, its definition requires some care: the states with the order parameters $+M_{0}$ and $-M_{0}$ are orthogonal and their Hilbert spaces are unitarily inequivalent. An infinitesimal external field can pin one of these degenerate vacuua. Thus, the broken-symmetry state has an effective  classical description in which the environmental degrees of freedom are relaxed around a specific state of the qubit. The corresponding uncertainty is zero and so is the von Neumann entropy. In the disordered state, by contrast, infinitesimal field has only infinitesimal effect. Right at the quantum critical point, the qubit is maximally entangled with the environment, as will be shown below. So the phase transition can be aptly described as a classical-to-quantum transition. There are a number of recent interesting proposals to observe this dissipative phase transition using a quantum dot. \cite{Borda:2006,LeHur:2005} The renormalization group flows for this transition are the same as those of the classical inverse-square Ising model in one dimension; $(\Delta/2\omega_{c})$ plays the role of the fugacity $y$ of the kinks in the Ising model, while $\alpha$ maps onto the inverse temperature. Although the linearized renormalization group equations are identical to the Kosterlitz-Thouless (KT) transition of the two-dimensional $XY$ model, the physics of the ordered phase is entirely different because of the existence of a local order parameter.

In the ground state of Eq.~\ref{hamiltonian}, the reduced density matrix of the spin degree of freedom is determined by the expectation values $\langle \sigma_x \rangle$ and $\langle \sigma_z \rangle$:
\begin{equation}
\rho_A=\frac{1}{2} \left( \begin{array}{cc}
                                             1+\langle \sigma_z \rangle & \langle \sigma_x \rangle \\
                                             \langle \sigma_x \rangle & 1-\langle \sigma_z \rangle
                                           \end{array} \right)
\end{equation}
After diagonalizing $\rho_A$, we can easily compute the ground state von Neumann entropy.
Hence the behavior of the entanglement at the QPT follows directly from the behavior of $\langle \sigma_x \rangle$ and $\langle \sigma_z \rangle$.
Since the order parameter $\langle \sigma_z \rangle$ is discontinuous at the transition, the von Neumann entropy also jumps by an amount $\Delta S$.  As shown in the Supplementary Information, the magnitude of this jump is, to leading order in $y$,
\begin{equation}
\Delta S =  \ln 2+(y/2) \ln y
\end{equation}
In the limit of vanishing $y$, the system goes from being unentangled ($\langle \sigma_x \rangle=0$, $\langle \sigma_z \rangle =1$) to being maximally entangled ($\langle \sigma_x \rangle=\langle \sigma_z \rangle =0$) as it enters the disordered state.  Note that this result depends crucially on a proper treatment of the broken symmetry---without the jump in the order parameter, $S$ would be continuous through the transition.  We also emphasize that the von Neumann entropy is discontinuous (with a singular derivative) even though the correlation length in imaginary time diverges at the transition.  Because this divergence takes the form of an essential singularity, it does not leave a strong signature in other quantities at the critical point.

In addition to the QPT discussed above, the spin-boson model also undergoes a dynamic crossover at $\alpha=1/2$, below which the spin can exhibit damped coherent oscillations on short time scales.  For $\alpha>1/2$ the dynamics of the disordered phase is characterized by an overdamped decay.  If the central claim of this paper is correct, the von Neumann entropy should be analytic at $\alpha=1/2$ because no phase transition occurs at this point.  This appears to be consistent with numerical renormalization group calculations of the anisotropic Kondo problem. \cite{Costi:2003}

We now turn to a simpler but equally important example, the damped harmonic oscillator.  Consider a single harmonic oscillator (momentum $p$, position $x$, mass $M$, and frequency $\omega_0$) coupled to an environment that also consists of harmonic oscillators. Again we examine the case of Ohmic dissipation, where the spectral function (defined as in Eq.~\ref{spectraldens}) is $J(\omega)=\eta\omega$ for small $\omega$ and zero above the cutoff $\omega_c$.  The ground state expectation values of $x^2$ and $p^2$ are given by \cite{Weiss:1999}
\begin{equation}
\begin{array}{lcl}
\langle x^2 \rangle & = & \frac{\hbar}{2M\omega_0} f(\kappa) \\
\langle p^2 \rangle & = & \frac{\hbar M \omega_0}{2}(1-2\kappa^2)f(\kappa)+\frac{2\hbar M \omega_0}{\pi} \, \kappa \ln(\frac{\omega_c}{\omega_0})
\end{array}
\end{equation}
where $\kappa=\eta/2M \omega_0$ is the friction coefficient and 
\begin{equation}
f(\kappa)=\frac{1}{\pi \sqrt{\kappa^2-1}} \ln \left( \frac{\kappa+\sqrt{\kappa^2-1}}{\kappa-\sqrt{\kappa^2-1}} \right)
\end{equation}
At $\kappa=1$ the system crosses over from damped oscillatiory to overdamped behavior ($ \sqrt{\kappa^2-1}$ is to be replaced by $ i \sqrt{1-\kappa^2}$), just as the spin-boson model does at $\alpha=1/2$.  The function
$f(\kappa)$ is real for all $\kappa>0$ and has identical power series expansion whether we approach $\kappa=1$ from the positive or the negative side. Thus it is analytic at $\kappa=1$.

At zero temperature, the normalized reduced density matrix for the damped harmonic oscillator has matrix elements \cite{Weiss:1999}
\begin{equation}
\langle x^{\prime} | \rho_A | x^{\prime\prime} \rangle = \sqrt{4b/\pi}\,e^{-a(x^{\prime}-x^{\prime\prime})^2-b(x^{\prime}+x^{\prime\prime})^2}
\end{equation}
where $a=\langle p^2 \rangle/2\hbar^2$ and $b=1/8\langle x^2 \rangle$.  To compute the von Neumann entropy, we first note \cite{Holzhey:1994}  that
\begin{equation}
\mathrm{Tr}(\rho_A \ln \rho_A)=\, \lim_{n\to1}\,\frac{\partial}{\partial n} \int dx^{\prime} \langle x^{\prime} | \rho_A^n | x^{\prime} \rangle
\end{equation}  
The subsequent derivation of $S$ is reproduced in the Supplementary Information; the result is
\begin{equation}
S =-\frac{1}{2}\ln \left( \frac{4b}{a-b} \right) +\frac{1}{2} \sqrt{\frac{a}{b}}\,\ln \left( \frac{\sqrt{a}+\sqrt{b}}{\sqrt{a}-\sqrt{b}} \right)
\label{dhoentropy}
\end{equation}
Note that $S\to 0$ as $b\to a$, which corresponds to the minimum uncertainty $\sqrt{\langle x^{2}\rangle\langle p^{2}\rangle}=\hbar/2$. The uncertainty relation is satisfied only for $b\le a$. We have verified that $S$ is analytic at $\kappa=1$ (see Supplementary Information)---this follows from the analyticity of the function $f(\kappa)$.

Despite the popularity of field-theoretic perturbative methods involving Feynman diagrams, some of the major breakthroughs in condensed matter physics have come from formulating wave functions that reflect the unique correlations of a quantum system with infinitely many degrees of freedom. The BCS wave function is remarkable because it incorporates the broken global gauge symmetry. The Bijl-Dingle-Jastrow-Feynman wave function for liquid $\mathrm{^{4}He}$ contains the essential ingredients of interacting superfluid matter. Similarly, the Laughlin wave function explains not only the fractional quantum Hall effect, but also the fractionalization of charge. We believe that the non-analyticity of the von Neumann entropy is a useful criterion for all quantum critical points at which the character of the ground state wave function changes. It provides a unique insight into the internal correlations of a many-body wave function. The approach described in the present paper can be applied to a wide variety of QPTs for which the conventional energy criterion fails. In particular, it can be immediately applied to transitions between integer quantum Hall plateaus. \cite{Sondhi:1997}

\section{Supplementary equations, figures and notes}
\subsection{Derivation of $\Delta S$ for the spin-boson model}

To compute $\Delta S$, we need to know the expectation values $\langle \sigma_x \rangle$ and $\langle \sigma_z \rangle$ at criticality.  Adapting the results of Anderson and Yuval \cite{Anderson:1971}, we find that
 $\langle \sigma_z \rangle$ jumps by $\sqrt{1/\alpha}$ along the quantum critical line $2y=-(1-\alpha)$.  The expectation value of $\sigma_x$ is most easily obtained by differentiating the ground state energy:  $\langle \sigma_x \rangle=-2 \, ( \partial E/\partial \Delta)$.  Since the ground state energy maps onto the free energy of the inverse-square Ising model, we can make use of the free energy scaling relation \cite{Anderson:1971,Young:1981}
\begin{equation}
dF(l)=y^2(l) \, e^{-l} dl
\label{freeenergy}
\end{equation}
Here $F$ is the free energy in units of $\omega_{c}$, so we have the direct correspondence $\langle \sigma_x \rangle=-2 \, ( \partial E/\partial \Delta) \rightarrow -\partial F/\partial y$.

On the separatrix $2y=-(1-\alpha)$, the flow of the fugacity is given by $y(l)=y/(1+2yl)$.  Since $y\to0$ as $l\to\infty$, the renormalization group equations (valid for $y\ll1$) become asymptotically exact in this limit.  The fixed point at $l=\infty$ corresponds to the fully polarized state of the Ising chain, for which the free energy is zero.  So if we tune to criticality, we can obtain a closed-form solution for $F$ by integrating Eq.~\ref{freeenergy} all the way to $l=\infty$.  Differentiating with respect to $y$ then gives
\begin{equation}
\langle \sigma_x \rangle = \frac{1}{2} \left[ 1-\frac{1}{2y}+\frac{1}{(2y)^2} \, e^{1/2y} \int_{1/2y}^{\infty} dt \,\frac{e^{-t}}{t} \right]
\label{sigmax}
\end{equation}
In the limit $y \rightarrow 0$, Eq.~\ref{sigmax} assumes the form $\langle \sigma_x \rangle \approx 2y$, while the discontinuity in $\langle \sigma_z \rangle$ goes as $1-y$.  Using the reduced density matrix given in the text, it is easy to show that
\begin{equation}
\Delta S =  \ln 2+(y/2) \ln y+\mathcal{O}(y)
\end{equation}

\subsection{Derivation of $S$ for the damped harmonic oscillator}

\begin{figure}
\begin{center}
\includegraphics[scale=0.5]{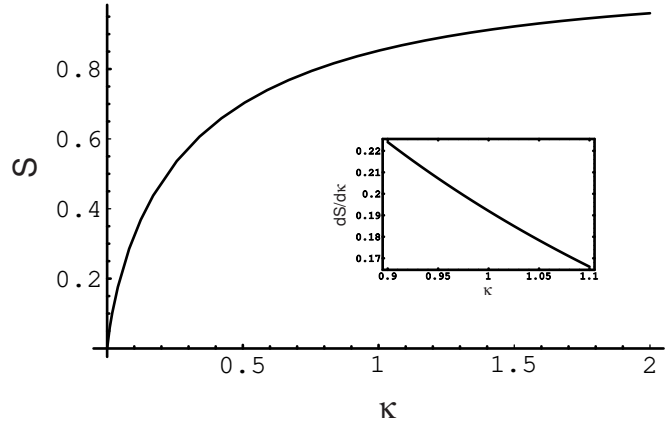}
\caption{Von Neumann entropy ($S$) of the damped harmonic oscillator with $\omega_c=100\, \omega_0$.  Both $S$ and its first derivative (inset) are perfectly smooth through the dynamic crossover at $\kappa=1$.  Higher-order derivatives are similarly well-behaved due to the analyticity of $f(\kappa)$ at $\kappa=1$.}
\label{fig:Sdho}
\end{center}
\end{figure}

As described in the text, the von Neumann entropy of the damped harmonic oscillator is given by
\begin{equation}
S=-\mathrm{Tr}(\rho_A \ln \rho_A)=\, -\lim_{n\to1}\,\frac{\partial}{\partial n} \int dx^{\prime} \langle x^{\prime} | \rho_A^n | x^{\prime} \rangle
\end{equation}
After inserting $n-1$ complete sets of position eigenstates, we can use the expression for the matrix elements to recast the integral as
\begin{equation}
\int dx^{\prime} \langle x^{\prime} | \rho_A^n | x^{\prime} \rangle = \left(\frac{4b}{\pi}\right)^{n/2} \int dx_1 \cdots dx_n \, e^{- x_i M_{ij} x_j}
\end{equation}
$\mathbf{M}$ is a tri-diagonal matrix with $M_{ii}=2(a+b)$ and $M_{i+1,i}=M_{i,i+1}=-(a-b)$, and therefore has eigenvalues $\gamma_m=2(a+b)-2(a-b)\cos (2\pi m/n)$, with $m=1,\ldots,n$.  So the von Neumann entropy becomes
\begin{equation}
S=-\,\lim_{n\to1}\,\frac{\partial}{\partial n} \left[ (4b)^{n/2} \left( \prod_{m=1}^{n} \gamma_m \right)^{-1/2} \right]
\label{entropy}
\end{equation}
For convenience, we write
\begin{equation}
\prod_{m=1}^{n} \gamma_m = 2^n (a-b)^n P
\label{det}
\end{equation}
where the product $P$ is defined as
\begin{equation}
P=\prod_{m=1}^{n} \left[\alpha-\cos\left(\frac{2\pi m}{n}\right)\right] ; \quad \alpha=\frac{a+b}{a-b}
\label{P}
\end{equation}
\begin{figure}
\begin{center}
\includegraphics[scale=0.5]{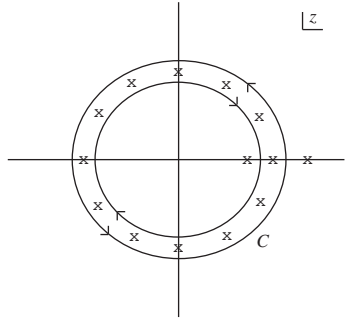}
\caption{The contour $C$ encloses the poles on the unit circle at $z_m=e^{i2\pi m/n}$.  With $\alpha>1$ ($b<a$), the additional poles at $z_{\pm}=\alpha \pm \sqrt{\alpha^2-1}$ are located on the real axis.}
\label{fig:contour}
\end{center}
\end{figure}
Our main task is to evaluate this product by performing a suitable integral transform.  Note that
\begin{equation}
\frac{\partial \ln P}{\partial \alpha} = \sum_{m=1}^{n} \frac{1}{\alpha-\cos(\frac{2\pi m}{n})}=\frac{1}{2\pi i}\oint_{C} dz \, \frac{1}{\alpha-\frac{1}{2}(z+\frac{1}{z})} \, \frac{nz^{n-1}}{z^n-1}
\end{equation}
provided the contour $C$ encloses the poles at $z_m=e^{i2\pi m/n}$ (Fig.~\ref{fig:contour}).  Deforming the contour to pick up the additional poles at $z_{\pm}=\alpha \pm \sqrt{\alpha^2-1}$, we obtain
\begin{equation}
\frac{\partial \ln P}{\partial\alpha}=\frac{2n}{z_{+}-z_{-}}\left( \frac{z_{+}^{n}}{z_{+}^{n}-1}-\frac{z_{-}^{n}}{z_{-}^{n}-1}\right)=\frac{2n}{z_{+}-z_{-}} \, \frac{z_{+}^{n}+1}{z_{+}^{n}-1}
\end{equation}
where the second equality follows from $z_{+}z_{-}=1$.  Integrating this result over $\alpha$ yields a closed-form expression for $P$.  The integral is most easily evaluated by transforming to the variable $z_{+}$:
\renewcommand{\arraystretch}{2}
\begin{equation}
\begin{array}{lcl}
\ln P & = & {\displaystyle \int dz_{+} \, \frac{n}{z_{+}} \, \frac{z_{+}^{n}+1}{z_{+}^{n}-1}} \\
         & = & {\displaystyle \ln \left( z_{+}^{n} + \frac{1}{z_{+}^{n}} -2 \right) + \rm{const.}}
\end{array}
\end{equation}
This is consistent with Eq.~\ref{P} if we set the integration constant equal to $-n \ln 2$.  In terms of the original parameters $a$ and $b$, the product is given by
\begin{equation}
P=\frac{1}{2^n} \left[ \left( \frac{\sqrt{a}+\sqrt{b}}{\sqrt{a}-\sqrt{b}} \right)^n+\left( \frac{\sqrt{a}-\sqrt{b}}{\sqrt{a}+\sqrt{b}} \right)^n-2 \right]
\end{equation}
and the rest of the calculation follows trivially from Eqs.~{\ref{entropy} and \ref{det}}.  The von Neumann entropy is
\begin{equation}
S =-\frac{1}{2}\ln \left( \frac{4b}{a-b} \right) +\frac{1}{2} \sqrt{\frac{a}{b}}\,\ln \left( \frac{\sqrt{a}+\sqrt{b}}{\sqrt{a}-\sqrt{b}} \right)
\end{equation}
where the ratio $b/a$ depends on the friction coefficient $\kappa$ and the ultraviolet cutoff $\omega_c$ (recall that the uncertainty principle requires $b\le a$).  $S$ is an analytic function of $\kappa$ as long as $b/a$ is analytic---which, in turn, is guaranteed by the analyticity of the function $f(\kappa)$.   The plots in Fig.~\ref{fig:Sdho} provide further proof of this point.

\vspace{0.5cm}

We thank J. Rudnick for helpful discussions. This work was supported by the NSF under grant:  DMR-0411931.


\begin{thebibliography}{99}
\expandafter\ifx\csname url\endcsname\relax
  \def\url#1{\texttt{#1}}\fi
\expandafter\ifx\csname urlprefix\endcsname\relax\def\urlprefix{URL }\fi
\providecommand{\bibinfo}[2]{#2}
\providecommand{\eprint}[2][]{\url{#2}}

\bibitem{Yang:1952}
\bibinfo{author}{Yang, C.~N.} \& \bibinfo{author}{Lee, T.~D.}
\newblock \bibinfo{title}{Statistical theory of equations of state and phase
  transitions .1. \textrm{T}heory of condensation}.
\newblock \emph{\bibinfo{journal}{Phys. Rev.}} \textbf{\bibinfo{volume}{87}},
  \bibinfo{pages}{404--409} (\bibinfo{year}{1952}).

\bibitem{Sachdev:1999}
\bibinfo{author}{Sachdev, S.}
\newblock \emph{\bibinfo{title}{Quantum phase transitions}}
  (\bibinfo{publisher}{Cambridge University Press},
  \bibinfo{address}{Cambridge}, \bibinfo{year}{1999}).

\bibitem{Edwards:1971}
\bibinfo{author}{Edwards, J.~T.} \& \bibinfo{author}{Thouless, D.~J.}
\newblock \bibinfo{title}{Regularity of density of states in \textrm{A}ndersons
  localized electron model}.
\newblock \emph{\bibinfo{journal}{J. Phys. \textrm{C}}}
  \textbf{\bibinfo{volume}{4}}, \bibinfo{pages}{453} (\bibinfo{year}{1971}).

\bibitem{Bennett:1996}
\bibinfo{author}{Bennett, C.~H.}, \bibinfo{author}{Bernstein, H.~J.},
  \bibinfo{author}{Popescu, S.} \& \bibinfo{author}{Schumacher, B.}
\newblock \bibinfo{title}{Concentrating partial entanglement by local
  operations}.
\newblock \emph{\bibinfo{journal}{Phys. Rev. \textrm{A}}}
  \textbf{\bibinfo{volume}{53}}, \bibinfo{pages}{2046--2052}
  (\bibinfo{year}{1996}).

\bibitem{Abrahams:1979}
\bibinfo{author}{Abrahams, E.}, \bibinfo{author}{Anderson, P.~W.},
  \bibinfo{author}{Licciardello, D.~C.} \& \bibinfo{author}{Ramakrishnan,
  T.~V.}
\newblock \bibinfo{title}{Scaling theory of localization: Absence of quantum
  diffusion in two dimensions}.
\newblock \emph{\bibinfo{journal}{Phys. Rev. Lett.}}
  \textbf{\bibinfo{volume}{42}}, \bibinfo{pages}{673--676}
  (\bibinfo{year}{1979}).

\bibitem{Chakravarty:1982}
\bibinfo{author}{Chakravarty, S.}
\newblock \bibinfo{title}{Quantum fluctuations in the tunneling between
  superconductors}.
\newblock \emph{\bibinfo{journal}{Phys. Rev. Lett.}}
  \textbf{\bibinfo{volume}{49}}, \bibinfo{pages}{681--684}
  (\bibinfo{year}{1982}).

\bibitem{Osterloch:2002}
\bibinfo{author}{Osterloh, A.}, \bibinfo{author}{Amico, L.},
  \bibinfo{author}{Falci, G.} \& \bibinfo{author}{Fazio, R.}
\newblock \bibinfo{title}{Scaling of entanglement close to a quantum phase
  transition}.
\newblock \emph{\bibinfo{journal}{Nature}} \textbf{\bibinfo{volume}{416}},
  \bibinfo{pages}{608--610} (\bibinfo{year}{2002}).

\bibitem{Osborne:2002}
\bibinfo{author}{Osborne, T.~J.} \& \bibinfo{author}{Nielsen, M.~A.}
\newblock \bibinfo{title}{Entanglement in a simple quantum phase transition}.
\newblock \emph{\bibinfo{journal}{Phys. Rev. \textrm{A}}}
  \textbf{\bibinfo{volume}{66}}, \bibinfo{pages}{032110}
  (\bibinfo{year}{2002}).

\bibitem{Refael:2004}
\bibinfo{author}{Refael, G.} \& \bibinfo{author}{Moore, J.~E.}
\newblock \bibinfo{title}{Entanglement entropy of random quantum critical
  points in one dimension}.
\newblock \emph{\bibinfo{journal}{Phys. Rev. Lett.}}
  \textbf{\bibinfo{volume}{93}}, \bibinfo{pages}{260602}
  (\bibinfo{year}{2004}).

\bibitem{Vidal:2003}
\bibinfo{author}{Vidal, G.}, \bibinfo{author}{Latorre, J.~I.},
  \bibinfo{author}{Rico, E.} \& \bibinfo{author}{Kitaev, A.}
\newblock \bibinfo{title}{Entanglement in quantum critical phenomena}.
\newblock \emph{\bibinfo{journal}{Phys. Rev. Lett.}}
  \textbf{\bibinfo{volume}{90}}, \bibinfo{pages}{227902}
  (\bibinfo{year}{2003}).

\bibitem{Bekenstein:1973}
\bibinfo{author}{Bekenstein, J.~D.}
\newblock \bibinfo{title}{Black holes and entropy}.
\newblock \emph{\bibinfo{journal}{Phys. Rev. D}} \textbf{\bibinfo{volume}{7}},
  \bibinfo{pages}{2333--2346} (\bibinfo{year}{1973}).

\bibitem{Laughlin:1983}
\bibinfo{author}{Laughlin, R.~B.}
\newblock \bibinfo{title}{Anomalous quantum hall-effect: An incompressible
  quantum fluid with fractionally charged excitations}.
\newblock \emph{\bibinfo{journal}{Phys. Rev. Lett.}}
  \textbf{\bibinfo{volume}{50}}, \bibinfo{pages}{1395--1398}
  (\bibinfo{year}{1983}).

\bibitem{Feenberg:1969}
\bibinfo{author}{Feenberg, E.}
\newblock \emph{\bibinfo{title}{Theory of quantum fluids}}
  (\bibinfo{publisher}{Academic Press}, \bibinfo{address}{New York},
  \bibinfo{year}{1969}).

\bibitem{Stauber:2006}
\bibinfo{author}{Stauber, T.} \& \bibinfo{author}{Guinea, F.}
\newblock \bibinfo{title}{Entanglement and dephasing of quantum dissipative
  systems}.
\newblock \emph{\bibinfo{journal}{Phys. Rev. A}} \textbf{\bibinfo{volume}{73}}
  (\bibinfo{year}{2006}).

\bibitem{Wegner:1980}
\bibinfo{author}{Wegner, F.}
\newblock \bibinfo{title}{Inverse participation ratio in $(2 +
  \epsilon)$-dimensions}.
\newblock \emph{\bibinfo{journal}{Z. Phys. \textrm{B}}}
  \textbf{\bibinfo{volume}{36}}, \bibinfo{pages}{209--214}
  (\bibinfo{year}{1980}).

\bibitem{Zanardi:2002}
\bibinfo{author}{Zanardi, P.}
\newblock \bibinfo{title}{Quantum entanglement in fermionic lattices}.
\newblock \emph{\bibinfo{journal}{Phys. Rev. \textrm{A}}}
  \textbf{\bibinfo{volume}{65}}, \bibinfo{pages}{042101}
  (\bibinfo{year}{2002}).

\bibitem{Calabrese:2004}
\bibinfo{author}{Calabrese, P.} \& \bibinfo{author}{Cardy, J.}
\newblock \bibinfo{title}{Entanglement entropy and quantum field theory}.
\newblock \emph{\bibinfo{journal}{Journal of Statistical Mechanics: Theory and
  Experiment}} \textbf{\bibinfo{volume}{2005}}, \bibinfo{pages}{P04010}
  (\bibinfo{year}{2005}).
\newblock \urlprefix\url{http://stacks.iop.org/1742-5468/2005/P04010}.

\bibitem{Hofstetter:1994}
\bibinfo{author}{Hofstetter, E.} \& \bibinfo{author}{Schreiber, M.}
\newblock \bibinfo{title}{Relation between energy-level statistics and
  phase-transition and its application to the \textrm{A}nderson model}.
\newblock \emph{\bibinfo{journal}{Phys. Rev. \textrm{B}}}
  \textbf{\bibinfo{volume}{49}}, \bibinfo{pages}{14726--14729}
  (\bibinfo{year}{1994}).

\bibitem{Parshin:1999}
\bibinfo{author}{Parshin, D.~A.} \& \bibinfo{author}{Schober, H.~R.}
\newblock \bibinfo{title}{Distribution of fractal dimensions at the
  \textrm{A}nderson transition}.
\newblock \emph{\bibinfo{journal}{Phys. Rev. Lett.}}
  \textbf{\bibinfo{volume}{83}}, \bibinfo{pages}{4590--4593}
  (\bibinfo{year}{1999}).

\bibitem{Bhattacharjee:2001}
\bibinfo{author}{Bhattacharjee, S.~M.} \& \bibinfo{author}{Seno, F.}
\newblock \bibinfo{title}{A measure of data collapse for scaling}.
\newblock \emph{\bibinfo{journal}{J. Phys. \textrm{A}}}
  \textbf{\bibinfo{volume}{34}}, \bibinfo{pages}{6375--6380}
  (\bibinfo{year}{2001}).

\bibitem{Zurek:1993}
\bibinfo{author}{Zurek, W.~H.}, \bibinfo{author}{Habib, S.} \&
  \bibinfo{author}{Paz, J.~P.}
\newblock \bibinfo{title}{Coherent states via decoherence}.
\newblock \emph{\bibinfo{journal}{Phys. Rev. Lett.}}
  \textbf{\bibinfo{volume}{70}}, \bibinfo{pages}{1187--1190}
  (\bibinfo{year}{1993}).

\bibitem{Leggett:1987}
\bibinfo{author}{Leggett, A.~J.} \emph{et~al.}
\newblock \bibinfo{title}{Dynamics of the dissipative two-state system}.
\newblock \emph{\bibinfo{journal}{Rev. Mod. Phys.}}
  \textbf{\bibinfo{volume}{59}}, \bibinfo{pages}{1--85} (\bibinfo{year}{1987}).

\bibitem{Blume:1970}
\bibinfo{author}{Blume, M.}, \bibinfo{author}{Emery, V.~J.} \&
  \bibinfo{author}{Luther, A.}
\newblock \bibinfo{title}{Spin-boson systems: One-dimensional equivalents and
  \textrm{K}ondo problem}.
\newblock \emph{\bibinfo{journal}{Phys. Rev. Lett.}}
  \textbf{\bibinfo{volume}{25}}, \bibinfo{pages}{450--453}
  (\bibinfo{year}{1970}).

\bibitem{Borda:2006}
\bibinfo{author}{Borda, L.}, \bibinfo{author}{Zarand, G.} \&
  \bibinfo{author}{Goldhaber-Gordon, D.}
\newblock \bibinfo{title}{Dissipative quantum phase transition in a quantum
  dot}.
\newblock \emph{\bibinfo{journal}{http://arxiv.org/abs/cond-mat/0602019}}
  (\bibinfo{year}{2006}).

\bibitem{LeHur:2005}
\bibinfo{author}{Le~Hur, K.} \& \bibinfo{author}{Li, M.~R.}
\newblock \bibinfo{title}{Unification of electromagnetic noise and
  \textrm{L}uttinger liquid via a quantum dot}.
\newblock \emph{\bibinfo{journal}{Phys. Rev. \textrm{B}}}
  \textbf{\bibinfo{volume}{72}}, \bibinfo{pages}{073305}
  (\bibinfo{year}{2005}).

\bibitem{Costi:2003}
\bibinfo{author}{Costi, T.~A.} \& \bibinfo{author}{McKenzie, R.~H.}
\newblock \bibinfo{title}{Entanglement between a qubit and the environment in
  the spin-boson model}.
\newblock \emph{\bibinfo{journal}{Phys. Rev. \textrm{A}}}
  \textbf{\bibinfo{volume}{68}}, \bibinfo{pages}{034301}
  (\bibinfo{year}{2003}).

\bibitem{Weiss:1999}
\bibinfo{author}{Weiss, U.}
\newblock \emph{\bibinfo{title}{Quantum dissipative systems}}
  (\bibinfo{publisher}{World Scientific}, \bibinfo{address}{Singapore},
  \bibinfo{year}{1999}).

\bibitem{Holzhey:1994}
\bibinfo{author}{Holzhey, C.}, \bibinfo{author}{Larsen, F.} \&
  \bibinfo{author}{Wilczek, F.}
\newblock \bibinfo{title}{Geometric and renormalized entropy in conformal
  field-theory}.
\newblock \emph{\bibinfo{journal}{Nucl. Phys. \textrm{B}}}
  \textbf{\bibinfo{volume}{424}}, \bibinfo{pages}{443--467}
  (\bibinfo{year}{1994}).

\bibitem{Sondhi:1997}
\bibinfo{author}{Sondhi, S.~L.}, \bibinfo{author}{Girvin, S.~M.},
  \bibinfo{author}{Carini, J.~P.} \& \bibinfo{author}{Shahar, D.}
\newblock \bibinfo{title}{Continuous quantum phase transitions}.
\newblock \emph{\bibinfo{journal}{Rev. Mod. Phys.}}
  \textbf{\bibinfo{volume}{69}}, \bibinfo{pages}{315--333}
  (\bibinfo{year}{1997}).
  
\bibitem[30]{Anderson:1971} Anderson, P. W. and Yuval, G.  Some Numerical Results on \textrm{K}ondo Problem and Inverse Square One-Dimensional Ising Model.  J. Phys. C {\bf 4}, 607-620 (1971).

\bibitem[31]{Young:1981} Young, A. P. and Bohr, T.  Crossover in the Two-Dimensional \textrm{C}oulomb Gas.  J. Phys. \textrm{C} {\bf 14}, 2713-2721 (1981).

\end{thebibliography}
\end{document}